\documentclass[article]{aa}
\usepackage{txfonts}
\usepackage{epsfig} 
\usepackage{graphicx,rotating} 
\usepackage{amssymb} 
\usepackage{color} 
\usepackage{url} 
\usepackage{natbib}
\usepackage{multirow}
\bibpunct{(}{)}{;}{a}{}{,} 

\begin{document}


\title{Helioseismology of sunspots: how sensitive are travel times to the Wilson depression and to the subsurface magnetic field?}
\titlerunning{Helioseismology of sunspots}

\author{
         H. ~Schunker \inst{1}
  \and L. ~Gizon \inst{1,2}
  \and R.~H. ~Cameron \inst{1}
  \and A.~C. ~Birch \inst{1}
     }

\offprints{H. Schunker, \email{schunker@mps.mpg.de}}

\institute{Max-Planck-Institut f\"{u}r Sonnensystemforschung, 37191 Katlenburg-Lindau, Germany email: \texttt{[schunker;gizon;cameron;birch]@mps.mpg.de}
  \and Institut f\"{u}r Astrophysik, Georg-August-Universit\"{a}t, 37077 G\"{o}ttingen, Germany
}

\date{Received / Accepted }

\abstract{
In order to assess the ability of helioseismology to probe the subsurface structure and magnetic field of sunspots, we need to determine how helioseismic travel times depend on perturbations to sunspot models. 
Here we numerically simulate the propagation of f, p$_1$, and p$_2$ wave packets through magnetic sunspot models.
Among the models we considered, a $\pm$50~km change in the height of the Wilson depression and a change in the subsurface magnetic field geometry can both be detected above the observational noise level. 
We also find that the travel-time shifts due to changes in a sunspot model must be modeled by computing the effects of changing the reference sunspot model, and not by computing the effects of changing the subsurface structure in the quiet-Sun model.
For p$_1$ modes the latter is wrong by a factor of four.
In conclusion, numerical modeling of MHD wave propagation is an essential tool for the interpretation of the effects of sunspots on seismic waveforms.
}

\maketitle


\section{A numerical approach}\label{sec:intro}

Elucidating the subsurface processes of sunspot formation and evolution is important for understanding solar magnetism \citep{Moradi2010,RempelSchlich2011}. Helioseismology is the only available tool for imaging the subsurface structure, magnetic field and dynamics of sunspots \citep{GizonARAA2010}, all of which are largely unknown. 

Early approaches interpreted the seismic signature of sunspots in terms of small perturbations to a quiet-Sun reference model \citep{Fan1995,Kosovichev1996}. However, the effects of the surface magnetic field and associated structural changes cause large perturbations in the wave speed \citep[e.g.][]{Lindsey20051,GizonHB2006}. Recently, direct numerical simulations of MHD wave propagation through sunspot models have been used to capture the large effects of the magnetic field on the waves \citep{GizonARAA2010,Cameron2011}.

A study by \cite{Liang2013} measured the travel-time shifts caused by the sunspot in AR 9787. In an effort to understand and interpret these measurements we used numerical simulations.
We simulated the interaction of surface-gravity (f) and acoustic (p$_1$ and p$_2$) wave packets with several sunspot models. With spectroscopy the height of the Wilson depression in sunspots is not known to better than 50~km \citep{Solanki1993}. Here we investigate if seismic diagnostics can provide a stronger constraint. Further, we study the seismic consequences of two subsurface magnetic field geometries below sunspots: a monolithic flux tube versus a weaker, spread-out magnetic field. Finally, we explore the sensitivity of waves to sound-speed perturbations at different depths along the axis of a sunspot model.

These parametric studies enable us to test the assumptions of existing approximation methods used in the interpretation of sunspot seismology. By comparing with measured noise in seismic travel times we are able to draw conclusions about the detectability of different aspects of sunspot structure.

\section{Sunspot models}\label{sec:ssmodels}

\subsection{Background quiet-Sun model (QS)}\label{sec:qs}
Our sunspot models are embedded in the background solar model CSM\_A described by \cite{Schunker2011}. The CSM\_A model is based on Model~S from \citet{JCD1996}, but was modified to be convectively stable so that linear waves can be propagated using the \textsf{SLiM} wave simulation code \citep{Cameron2008}. An indication of the sensitivity of the modes of oscillation in CSM\_A as a function of  height, $z$, is given by their kinetic energy density (see Fig.~\ref{fig:mods}). Throughout this paper we define $z$ as height measured upward from the bottom of the photosphere.

\subsection{Reference sunspot model (RS)}\label{sec:rs}
\begin{center}
\begin{figure}
\vspace{-2.5cm}
\hspace{0.0cm}
\includegraphics[width=9cm]{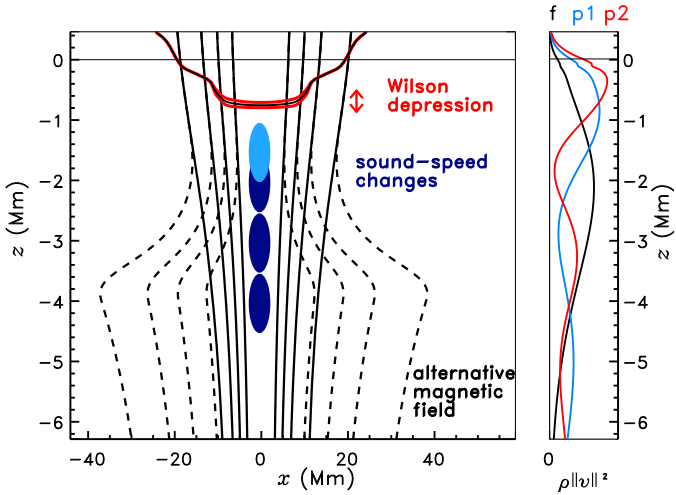} 
\caption{
Reference sunspot model and perturbations to it. The magnetic field lines of the reference model are shown by the black curves (Sect.~2.1.2). The dashed black curves are the magnetic field lines of the alternative magnetic field model (Sect.~2.1.4). Two Wilson depressions modified by $\pm 50$~km have been considered (red curves, Sect.~2.1.3).
The ellipses along the axis of the sunspot indicate the locations of the various sound-speed perturbations (Sect.~2.1.5). The size of the ellipses corresponds to the full width at half maximum of the Gaussian perturbations. The right panel shows mode kinetic energy density, $\rho \| \mathbf{v} \|^2$, for the f mode at $kR_\odot=450$ (black), for p$_1$ at $kR_\odot=420$ (blue), and for p$_2$ at $kR_\odot=390$ (red), where $\rho$ is density and $\mathbf{v}$ is the velocity eigenfunction for quiet-Sun model CSM\_A. Mode kinetic energy density is a useful indicator for the vertical sensitivity of the modes.
}
\label{fig:mods}
\end{figure}
\end{center}

\cite{Cameron2011} constructed a parametric sunspot model that causes a seismic response consistent with that observed for the sunspot in active region NOAA 9787 \citep[][, Paper I]{Liang2013}. 
The vector magnetic field of the sunspot model is cylindrically symmetric, i.e. it depends only  on the horizontal distance from the axis of the sunspot, $\varpi=(x^2+y^2)^{1/2}$, and on height, $z$. The model of \cite{Cameron2011}  is a self-similar and monolithic magnetic field model specified by the strength of the field on the axis of the sunspot ($\varpi=0$) as a function of height.  Here we adopt  the same sunspot model with the parameters given in Table~1 of \cite{Cameron2011}, except that the parameters controlling the inclination of the magnetic field are modified slightly ($\alpha_1=1.5$~Mm and $\alpha_2=9.2$~Mm) so that the magnetic field inclination  at the umbra-penumbra boundary ($50^\circ$) better matches observed field inclinations in sunspots \citep{Schunker2008}. The solid black curves in Fig.~\ref{fig:mods} show a sample of the magnetic field lines. Throughout this paper we refer to this model as the reference sunspot model abbreviated as `RS'.

\subsection{Changes to the Wilson depression}\label{sec:wd}
The Wilson depression is a physical depression of the surface (optical depth unity) caused by strong magnetic fields in the umbra of a sunspot \citep{LoughheadBray1958}.  The Wilson depression of our reference sunspot model is at  $z=-550$~km in the center of the umbra and at $z=-300$~km in the middle of the penumbra as in \citet{Cameron2011}.
 
In order to assess the ability of helioseismology to measure the height of the Wilson depression,  we consider two additional sunspot models in which the Wilson depression is shifted by $\pm50$~km with respect to the reference sunspot model. In these modified models, we translate the density, pressure, and sound-speed profiles in height by $\pm 50$~km, while keeping the magnetic field unchanged (models are not in magnetohydrostatic equilibrium).

\subsection{Alternative magnetic field model}\label{sec:dc}
The monolithic reference sunspot model described in Sect.~\ref{sec:rs} is only one possibility among many. We considered a sunspot model with an alternative magnetic field below $z=-1$~Mm, which fans out as a function of depth. This alternative model is motivated by, e.g., the dynamical disconnection model of \cite{SchusslerRempel2005} whereby the flux tube is disrupted by convective motions at depths greater than ~2 Mm. In the alternative magnetic sunspot model, we keep the sound speed, pressure and density profiles unchanged from the reference sunspot model. 

More precisely, the alternative magnetic field is modified so that the magnetic field becomes weaker below $z_\mathrm{t}=-1$~Mm (above, it is identical to the reference sunspot model), smoothly transitioning to weak field strengths below $z_\mathrm{b}=-4$~Mm. Along the axis of the sunspot, the $z$-component of the magnetic field, $\tilde{B}_z$, is obtained from the reference field, $B_z$, as follows:  
\begin{equation}
   \tilde{B}_z(z, \varpi=0) = B_z(z, \varpi=0) 
   \left\{
        \begin{array}{ll}
      1 & \textrm{for} \, z \ge z_\mathrm{t} \, , \\
      \Pi(z) & \textrm{for} \, z_\mathrm{b} \le z \le z_\mathrm{t} \, , \\
      0.1 & \textrm{for} \, z \le z_\mathrm{b} .
     \end{array}
     \right.
\label{eqn:dc}
\end{equation}    	
where $\Pi(z)$ is the fifth order polynomial in $z$ such that $\tilde{B}_z(z)$ is continuous and has continuous first- and second-order derivatives, i.e. $\Pi(z)=1 - 0.9Z^3(6 Z^2 - 15 Z + 10)$ where $Z=(z - z_\mathrm{t})/(z_\mathrm{b} - z_\mathrm{t})$. The same field lines as for the reference sunspot model are shown for the alternative magnetic field model in Fig.~\ref{fig:mods}.

\subsection{Subsurface sound-speed perturbations} \label{sec:cs}
We also consider a set of sunspot models in which we modify the subsurface sound speed.  In these models, the sound speed is given by
\begin{equation}
\tilde{c}(\varpi,z) = c(\varpi,z) + \Delta c(\varpi,z)
\label{eqn:cs}
 \end{equation}
where $c$ is the sound speed in the reference sunspot model and  
\begin{equation}
\Delta c (\varpi,z)  = c(\varpi,z) A \exp\left[-\frac{\varpi^2}{2 \sigma_\varpi^2} - \frac{(z-h)^2}{2 \sigma_z^2}\right]
\label{eqn:deltacs}
\end{equation}
is a 3D Gaussian perturbation to the sunspot model. In the above expression, the relative amplitude of the perturbation is $A$, the height of the centre of the perturbation is $h$, and the width of the perturbations are  fixed to $\sigma_z=0.42$~Mm, $\sigma_\varpi=2.12$~Mm.  
For one experiment (Sect.~\ref{sec:sub}) we use $A=0.1$ and $h=-2,-3$ and $-4$~Mm. 
For a second experiment (Sect.~\ref{sec:sensitivity}) we use  $A=-0.01$ to $0.5$ and $h=-1.5$~Mm.


\section{Numerical simulations of wave propagation}\label{sec:sim}

We used the \textsf{SLiM} code  \citep{Cameron2007,Cameron2011}  to simulate the propagation of wave packets through the sunspot models described above. We use a simulation domain that covers $145.77$~Mm in each of the horizontal directions and extends from $-25$~Mm below to $2.5$~Mm above the bottom of the photosphere. The depth of the simulation domain is sufficient to  study acoustic modes with radial orders $n\le4$. The spatial resolutions are $0.025$~Mm per pixel horizontally and $0.73$~Mm per pixel vertically. The time resolution is $0.13$~s. We did a resolution study which showed that for the waves studied here the numerical solutions are converged.

We carry out simulations for individual wave packets with modes of fixed radial order, i.e. with either f, p$_1$, or p$_2$ modes. The initial conditions ($t=0$) for the simulations are specified by the displacement, $\boldsymbol{\xi}(x,y,z)$, and the velocity,$\mathbf{v}(x,y,z)$. This requires specifying the wave amplitude distribution, $A_k$, as a function of wavenumber, $k$ \citep[see][]{Cameron2008,Cameron2011}. For each wave packet $A_k$ is chosen in such a way that the vertical velocity in the simulation ($v_z$) is directly comparable to an observed cross-covariance function. This comparison is justified because the cross-covariance is closely related to the  Green's function in the far field \citep[e.g.,][]{GizonARAA2010}.

The wave packets in our simulations are comparable to the Solar and Heliospheric Observatory's Michelson Doppler Imager  (SOHO/MDI) observed cross-covariance functions presented by \citet{Liang2013}. The peak of wave power occurs at $kR_\odot=450$ and $\nu=2.1$~mHz for the f modes, at $kR_\odot=420$ and $\nu=2.5$~mHz for the p$_1$ modes, and at $kR_\odot=390$ and $\nu=3.1$~mHz for the p$_2$ modes. At $t=0$, each wave packet peaks at $x=-43.73$~Mm (at the left edge of Fig.~\ref{fig:ttmap}), from where they propagate in the $+x$ direction (to the right) and interact with the sunspot centred at $x=y=0$~Mm.

\section{Travel times}

\subsection{Measuring travel times}\label{sec:tt}
Travel-time measurements are very useful to characterize the sensitivity of seismic waveforms to subsurface perturbations. In this paper we explore how changes to the reference sunspot model  affect travel times.  

We define the travel-time shift, $\Delta\tau$, as the time by which a sliding reference wavelet, $w$, must be shifted  to best match the seismic waveform $v_z(x,y,t)=v_z(x,y,z=0.2~\mathrm{Mm},t)$. Following \cite{Gizon2002} and \cite{Liang2013} $\Delta \tau$ is the time, $\tau$, that  maximises the function
\begin{equation}
F(\tau) = \frac{ \int v_z(x,y,t)    \, w(x,y,t-\tau) \, \mathrm{d}t}{ \int | w(x,y,t)|^2 \, \mathrm{d}t }.
\label{eqn:f}
\end{equation}
In practice, the reference wavelet $w$ was chosen to be  $v_z$ from either the quiet-Sun simulation or the reference sunspot simulation. To denote these two kinds of  measurements (which give similar results, see Sect~\ref{sec:sensitivity}), we use the notations $\Delta_{QS}\tau$  and $\Delta_{RS} \tau$ respectively. Furthermore, we use the notation $\Delta \tau(\mathrm{MODEL})$ to indicate the background model through which the seismic waveform is propagated, where `MODEL' is, e.g., `RS', `RS+pert', or `QS+pert'. The time integrals in Eq.~\ref{eqn:f} are approximated by summations with a time discretisation of $6$~s. With this time discretisation travel times are accurate to $0.01$~s, which is sufficient for our purpose.

\subsection{Estimating travel-time noise}\label{sec:obs}

While the numerical simulations performed in this paper are used to estimate the expectation value of a travel-time measurement (the signal), we also need to estimate the noise level in order to determine if a signal is detectable.

\begin{center}
\begin{figure}
\vspace{-1cm}
\includegraphics[width=8.5cm]{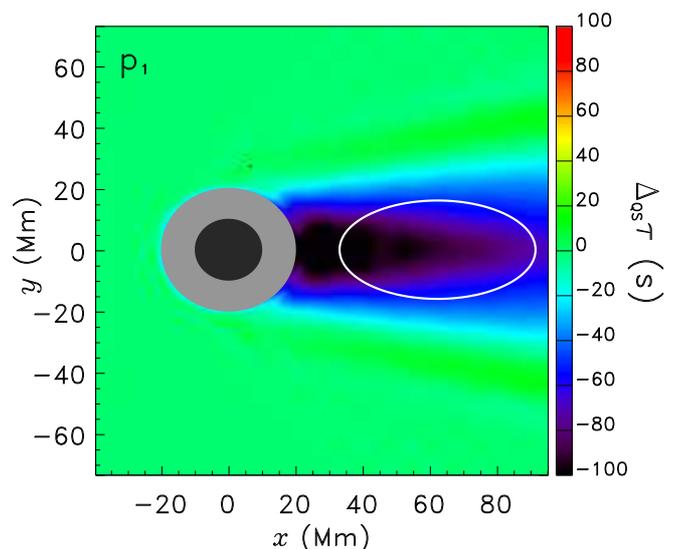}
\vspace{-0.5cm}
\caption{
Spatial map of travel-time shifts $\Delta_\mathrm{QS} \tau$ for the reference sunspot model.
The SLiM code numerically computes the waveform as a function of time through the reference sunspot model.
The color bar indicates  the travel-time shift  in seconds.
The ellipse encloses the region for which the spatial average of the travel-time shift (denoted with symbols $\langle \rangle$) is computed. 
The grey (black) circle indicates the penumbra (umbra) of the sunspot.
} 
\label{fig:ttmap}
\vspace{-0.5cm}
\end{figure}
\end{center}

We use the travel-time maps measured by \citet{Liang2013} covering seven independent days of SOHO/MDI observations of AR9787 \citep{Gizon2009,Moradi2010}. In order to decrease noise, we further average these travel-time maps over a region behind the sunspot where scattering phase shifts are large. This elliptical portion of the travel-time maps is shown in Fig.~\ref{fig:ttmap}. We use angle brackets $\langle \cdot \rangle$ to denote this spatial averaging  of the travel times.

For one day of observations, the spatially averaged travel-time shift introduced by the sunspot is $\langle \Delta_\mathrm{QS} \tau \rangle \simeq 75$~s for the p$_1$ modes.  In order to estimate the noise in these measurements, we compute $\langle \Delta_\mathrm{QS} \tau \rangle$  for each of the seven days of observations. We define the observed noise level as the standard error in the mean of the spatially averaged travel-time shifts. The observed noise level decreases with radial order, it is $3.8$~s for the f modes, $3.5$~s for p$_1$ modes and $1.6$~s for p$_2$ modes (Table~\ref{table:difftt}).

\section{Results}
We carried out two sets of numerical experiments. In the first set, we propagated f, p$_1$ and p$_2$ waves independently through different sunspot models to predict the travel-time shifts caused by various perturbations to the reference sunspot model, and to determine if they are above the observed noise level. 
In the second set of experiments we looked at the sensitivity of p$_1$ wave travel times to changes in the amplitude of the subsurface sound-speed. We considered sound-speed perturbations to both the quiet-Sun model and to the reference sunspot model. 
The results of these experiments are described in the sections below.


\subsection{Is it possible to distinguish between sunspot models?}\label{sec:sub}
One way to measure the effect of a change in the sunspot model on the waves is to look at the change in the travel-time shifts measured with respect to the waves in the quiet-Sun reference model, i.e. by computing the difference $\langle \Delta_\mathrm{QS} \tau(\mathrm{RS} + \mathrm{pert} ) \rangle- \langle \Delta_\mathrm{QS}  \tau(\mathrm{RS}) \rangle$. 

Table~\ref{table:difftt} shows these changes in travel-time shifts for the perturbations to the reference sunspot model  described in Sects.~\ref{sec:wd}, \ref{sec:dc} and \ref{sec:cs}.
For the case where the Wilson depression is moved deeper (shallower) by $50$~km, we find that the f wave packet travels slower (faster) and the p$_1$ and p$_2$ wave packets travel faster (slower).  Because the travel-time shifts for these two cases do not have the same amplitude, perturbation theory that is linear in the height of the Wilson depression will not fully capture this result. The physical cause of these effects is not clear. We note, however, that a change in the Wilson depression implies a change in the density, sound-speed and Alfv\'en speed which all affect the wave speed \citep{Hanasogeetal2012}. For the case of the p$_2$ wave packet the change in the travel-time shift is roughly twice the noise level, implying that the p$_2$ travel times are useful for constraining the height of the Wilson depression to an accuracy better than $50$~km (assuming the other sunspot parameters are well constrained).

Reducing the strength of the subsurface magnetic field causes a change in the travel-time shift of roughly  $3$~s for all wave packets. The physical process governing the interaction of the wave with the subsurface magnetic field is not clear. Recall that this change in the magnetic field is only below $z=-1$~Mm. As for the case of changes to the height of the Wilson depression, for the p$_2$ wave packet the change in the  travel-time shift is about twice that of the observed noise level. Assuming that the Wilson depression was well constrained (and all other sunspot parameters), the subsurface magnetic field may also be  constrained. 

Inserting subsurface sound-speed enhancements causes the p$_1$ and p$_2$ mode wave packets to travel faster than expected. The f wave packet is not sensitive to the subsurface sound-speed at all, which is also not surprising since it is a surface gravity wave. Unless the observed signal to noise can be increased for the p$_1$ and p$_2$ wave packets by a factor of at least $4$, sound-speed perturbations of this type will not be able to be detected with the spatially averaged travel-time shifts of these modes.

\begin{table}
\caption{The spatially averaged travel-time shift introduced by the perturbation to the sunspot model, $\langle   \Delta_\mathrm{QS}\tau(\mathrm{RS}+ \mathrm{pert}) \rangle  - \langle  \Delta_\mathrm{QS} \tau(\mathrm{RS}) \rangle$.}  
\vspace{-0.5cm}
\label{table:difftt} 
\centering 
\begin{tabular}{l c c c }
\hline  \hline
  Sunspot Perturbation & \multicolumn{3}{c}{Travel-time Shift (s)} \\
 &f  & p$_1$  & p$_2$ \\
\hline  
Wilson depression: & & & \\
50~km deeper   & $+0.4$ & $-2.2$ & $\mathbf{-3.8}$ \\
50~km shallower   & $-0.3$ & $+1.9$ & $\mathbf{+3.2}$ \\
\hline
Alternative magnetic field   & $-2.4$ & $-2.8$ & $\mathbf{-3.3}$ \\
\hline
Sound-speed perturbations: & & & \\
$A=0.1$, $h=-2$~Mm   & $\, \, 0.0$ & $-0.4$ & $-0.8$ \\
$A=0.1$, $h=-3$~Mm   & $\, \, 0.0$ & $-0.8$ & $-0.6$ \\
$A=0.1$, $h=-4$~Mm   & $\, \, 0.0$ & $-0.8$ & $-0.2$ \\
$A=0.5$, $h=-1.5$~Mm   & \ldots & $-1.2$ & \ldots \\
\hline
\end{tabular}

\tablefoot{
Negative (positive) indicates that the waves travel faster (slower) through the perturbed model than the reference sunspot model.
The observational noise level is given in the bottom row.
The systematic error introduced by the measurement procedure is $\approx 0.1$~s.
The noise level for seven days of SOHO/MDI observations for the f, p$_1$ and p$_2$ mode is $3.7$~s, $3.5$~s and $1.6$~s respectively.
The travel-time shifts in bold lie above the noise level.
}
\end{table}

\subsection{Can we avoid using numerical simulations?}\label{sec:sensitivity}

In the second experiment, we further investigate the sensitivity of travel-time shifts to changes in the subsurface sound-speed. \cite{Fan1995}, \cite{Kosovichev1996} and \cite{Birch2004} explored this problem using perturbation theory around  quiet-Sun reference models. Here we go beyond small amplitude perturbation theory and look at finite amplitude perturbations to the subsurface sound-speed. In addition, we look at the effects of identical sound-speed changes to a reference sunspot model and in the quiet-Sun model.

We insert Gaussian sound-speed perturbations of various amplitudes into the reference sunspot model (i.e. Eq.~\ref{eqn:deltacs} with $A=-0.01$ to $0.5$ and $h=-1.5$~Mm).  
In order to assess the significance of the reference model, we also insert identical sound-speed perturbations  into the background quiet-Sun model so that 
\begin{equation}
 \tilde{c_0} (\varpi,z)  = c_0(z) + \Delta c(\varpi,z),
\label{eqn:deltacsqs}
\end{equation}
where $c_0$ is the quiet-Sun sound speed, and $\Delta c(\varpi,z)$ is the perturbation to the reference sunspot model defined by Eq.~\ref{eqn:deltacs}.
We simulated the propagation of  p$_1$ wave packets through these models. 

\begin{center}
\begin{figure}
\includegraphics[width=8cm]{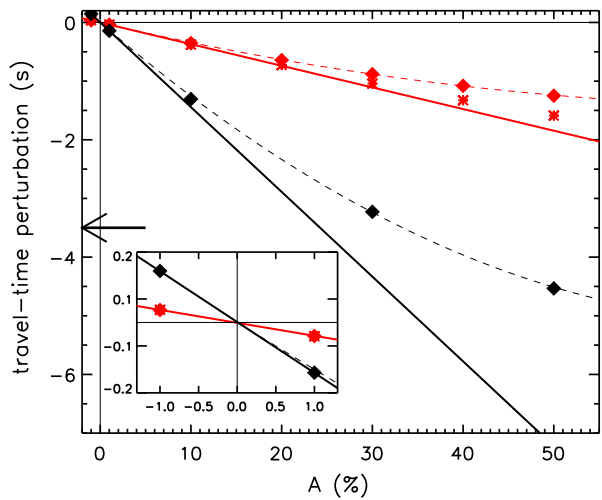}
\caption{
Travel-time perturbations versus sound-speed perturbation amplitudes $A$, for p$_1$ wave packets. The results from two SLiM numerical experiments are shown. 
In the first procedure, the red  diamonds show the travel-time perturbations resulting from sound-speed perturbations to a background model that includes the reference sunspot, $\langle \Delta_\mathrm{QS} \tau(\mathrm{RS}+\delta c)\rangle  -  \langle \Delta_\mathrm{QS} \tau(\mathrm{RS}) \rangle$. 
In the second procedure, the black  diamonds show the travel-time perturbations resulting from the same sound-speed perturbations to a quiet-sun background model, $\langle \Delta_\mathrm{QS} \tau(\mathrm{QS}+\delta c) \rangle$. 
In both experiments, the travel times (diamonds) are measured using sliding quiet-Sun waveforms.
For the first experiment, the red  asterisks show the results using a different measurement procedure, $\langle \Delta_\mathrm{RS} \tau(\mathrm{RS}+\delta c)\rangle$.
The solid lines show linear fits to the three smallest perturbations with A=-1\%, 0, and 1\%. 
The inset shows a zoom around the origin of the plot.
The dashed curves are parabolic fits.
The p$_1$ travel-time noise level for a 7 days  of observation is $3.5$~s isindicated by the arrow.
%
}
\label{fig:linear}
\end{figure}
\end{center}

Figure~\ref{fig:linear} shows the change in the spatially averaged travel-time shifts, $\langle \Delta_\mathrm{QS}  \tau (\mathrm{RS} + \delta c ) \rangle- \langle \Delta_\mathrm{QS}  \tau(\mathrm{RS}) \rangle$, as a function of the amplitude of the sound-speed perturbation. The ensuing changes to the travel-time shift are not linear with the amplitude of the sound-speed perturbation. The deviation from linearity for a sound-speed perturbation with a 10\% amplitude is 5\% -- a parabolic fit describes the change in the travel-time shifts with respect to sound-speed perturbation amplitude better. 
Notice that for all amplitudes, even up to 50\%, the change in the travel-time shift is well below the observed noise level. Thus, these subsurface sound-speed perturbations are not detectable using only the p$_1$ wave packet.

Another way to measure the travel-time shift due to the sound-speed perturbations is to directly compute the travel-time shift with respect to the reference sunspot model, i.e.
$\langle \Delta_\mathrm{RS} \tau(\mathrm{RS}+\delta c)\rangle$
(see Fig.~\ref{fig:linear}). This gives a different, but qualitatively similar, result to the previous method. This shows that the general behaviour of the change in the travel-time shifts with amplitude is not just a consequence of the measurement procedure.

Figure~\ref{fig:linear}  also shows the travel-time shifts, $\langle \Delta_\mathrm{QS}  \tau(\mathrm{QS} + \delta c)  \rangle$, caused by imposing the same sound-speed perturbations in the quiet-Sun model. In this case, the linear approximation  is incorrect by more than 10\% when the amplitude of the  sound-speed perturbation is 10\%.
More importantly, these travel-time shifts are roughly four times larger than the travel-time shifts caused by imposing the same sound-speed perturbation in the reference sunspot model. 
As a result, the quiet-Sun model cannot be used to predict the travel-time shifts caused by  sound-speed perturbations to a sunspot model because the sunspot  substantially changes the p$_1$ wave packet, and thus its sensitivity to subsurface perturbations. Therefore, it is necessary to use numerical simulations for a better understanding of the effects of a sunspot on the waves.

\section{Summary and prospects}\label{sec:conclusions}

We have shown that it should be possible to constrain the height of the Wilson depression to within $50$~km and to distinguish between the two models for the subsurface magnetic field structure (a monolithic and a weaker spreading field) using seven days of SOHO/MDI observations of one sunspot. 
Using only the spatial average of the travel-times shifts it is not possible to detect 10\% sound-speed changes with a length scale of 5~Mm at depths between $1.5$ and $4$~Mm with the same  observations.
We expect that including more information, e.g. the spatial distribution of the travel-time shifts, the amplitude of the cross-correlations, and higher order modes, will improve the ability of helioseismic measurements to constrain the subsurface structure of sunspots.

We have also demonstrated that the wave travel-time shifts caused by changes in sound speed to our sunspot model are different than the wave travel-time shifts caused by the same change in sound speed of the quiet-Sun model. 
This is due to the effect of the sunspot on the wavefield, and this effect is not captured by the traditional approximations of local helioseismology \citep{Fan1995,Kosovichev1996,Birch2004}. 
Computational helioseismology, however, provides a clear path forward. These simulations may also be used to test the magnetoghydrodynamic ray approximation \cite[e.g.][]{Cally2000,Schunker2006}.

Other avenues to explore are the use of vector magnetograms and intensity images to further constrain the surface properties of sunspots and the use of SDO-HMI Doppler observations \citep{HMI2012} for helioseismic measurements.

\begin{acknowledgements}
H.S. and L.G. acknowledge research funding by Deutsche Forschungsgemeinschaft (DFG) under grant SFB 963/1 ``Astrophysical flow instabilities and turbulence'' (Project A18,  WP ``Seismology of magnetic activity''). 
The German Data Center for SDO, funded by the German Aerospace Center (DLR), provided the IT infrastructure.
\end{acknowledgements}

\bibliographystyle{aa} 
\bibliography{master} 

\end{document}